\documentstyle[11pt]{article}

\input epsf.tex
\newcommand{\ba}{\begin{array}{c}}
\newcommand{\ea}{\end{array}}
\def\gz{\epsfxsize=.1cm\epsfbox{z0.eps}}
\def\gaa{\epsfxsize=.7cm\epsfbox{z11.eps}}
\def\gaab{\epsfxsize=1cm\epsfbox{z112.eps}}
\def\gbb{\epsfxsize=.7cm\epsfbox{z22.eps}}
\def\gcc{\epsfxsize=.7cm\epsfbox{z33.eps}}
\def\gaaab{\epsfxsize=.6cm\epsfbox[0 -50 252 253]{z1113.eps}}

\def\gaabb{\epsfxsize=1.2cm\epsfbox[0 -160 500 -124]{z1122.eps}}

\def\gbbb{\epsfxsize=.7cm\epsfbox[0 -60 288 192]{z222.eps}}
\def\Gbbb{\epsfxsize=.7cm\epsfbox{z222.eps}}
\def\gabc{\epsfxsize=1cm\epsfbox{z123.eps}}

\def\notwist{\ba\epsfxsize=3cm\epsfbox{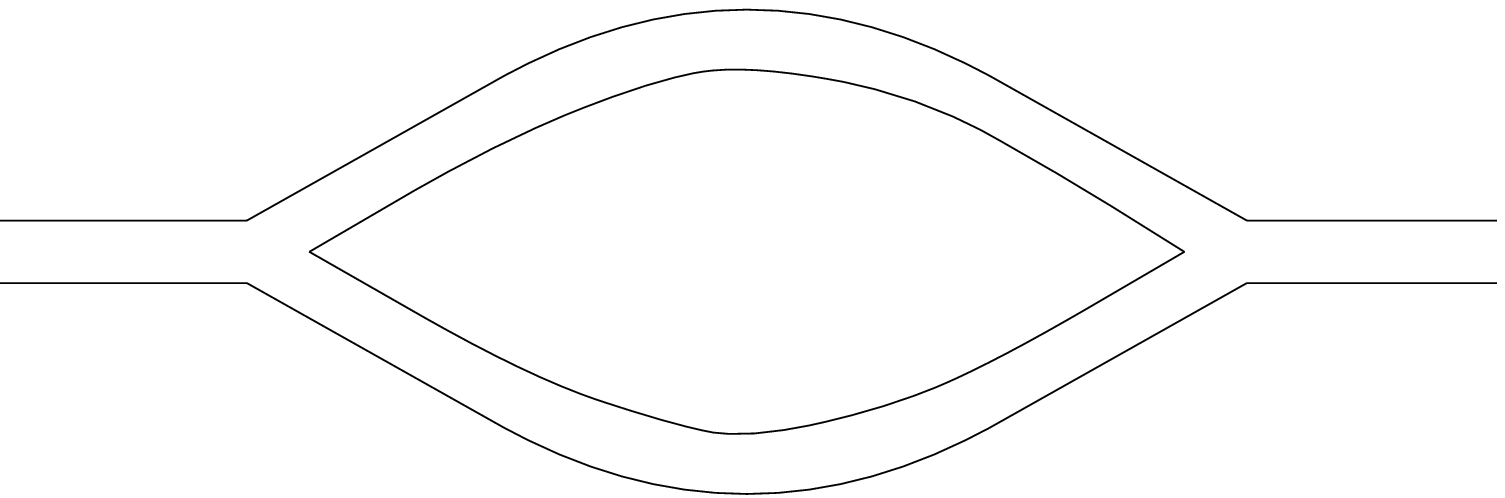}\ea}
\def\onetwist{\ba\epsfxsize=3cm\epsfbox{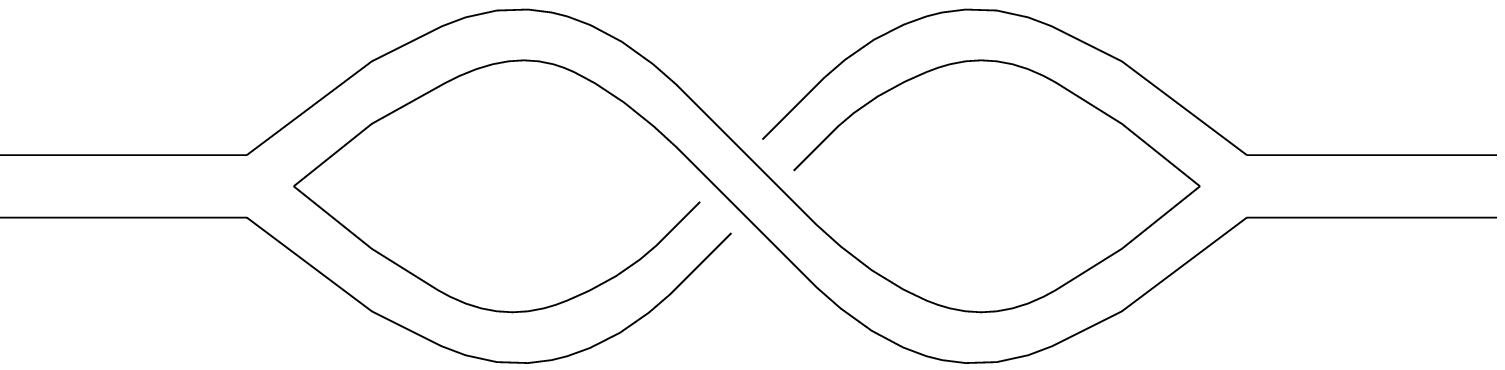}\ea}

\def\eqy{\epsfxsize=8cm\epsfbox{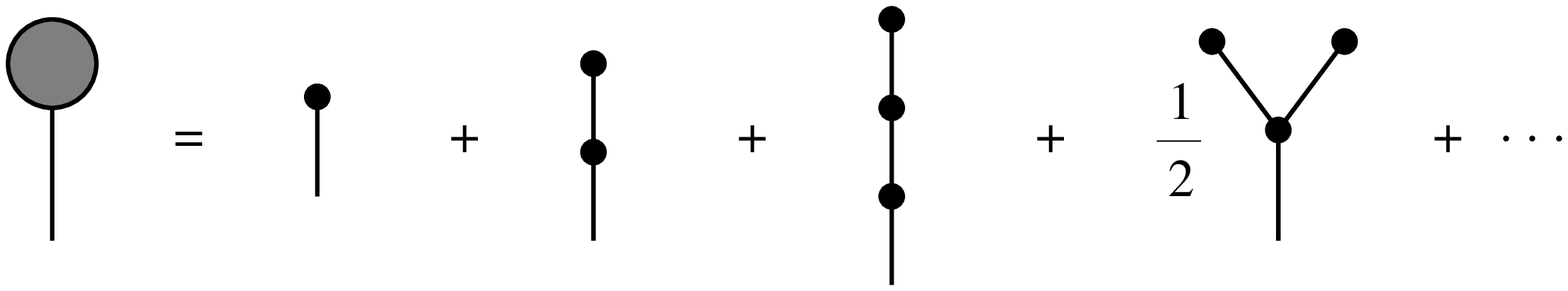}}
\def\vertnotwist{\epsfxsize=7cm\epsfbox{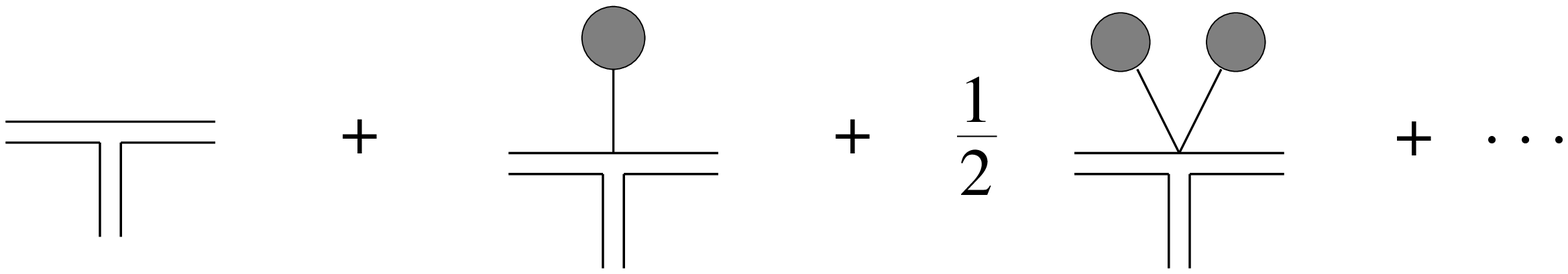}}
\def\vertonetwist{\epsfxsize=7cm\epsfbox{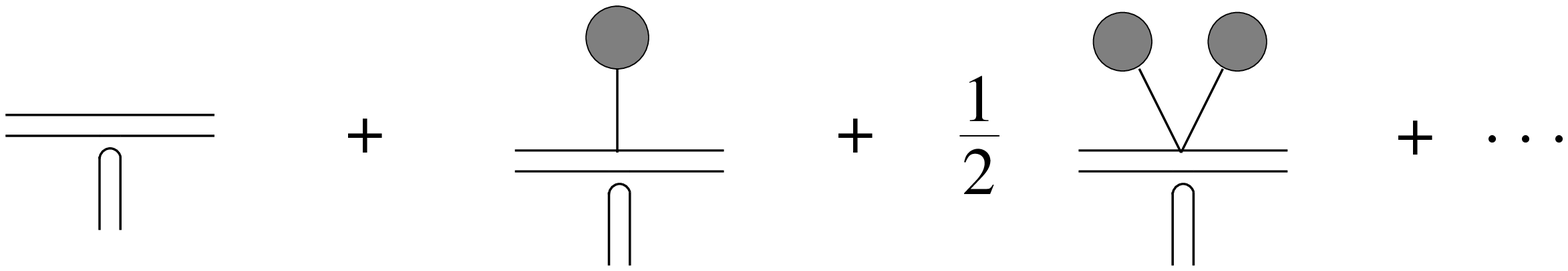}}
\def\exvert{\epsfxsize=2cm\epsfbox{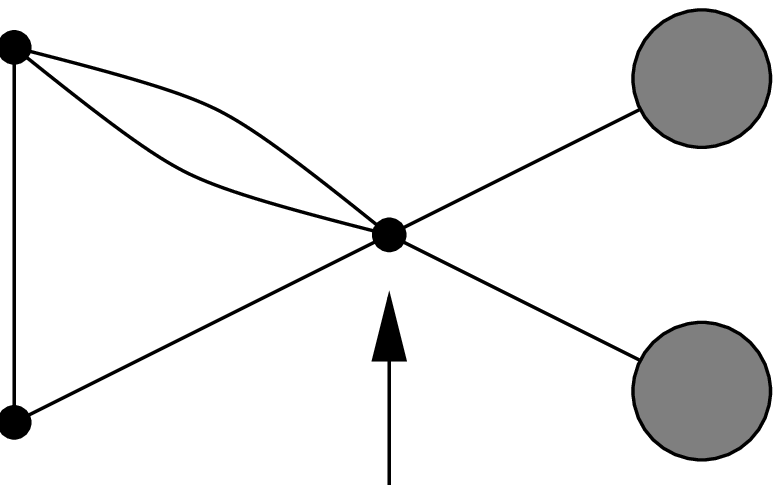}}
\def\notcubic{\epsfxsize=5cm\epsfbox{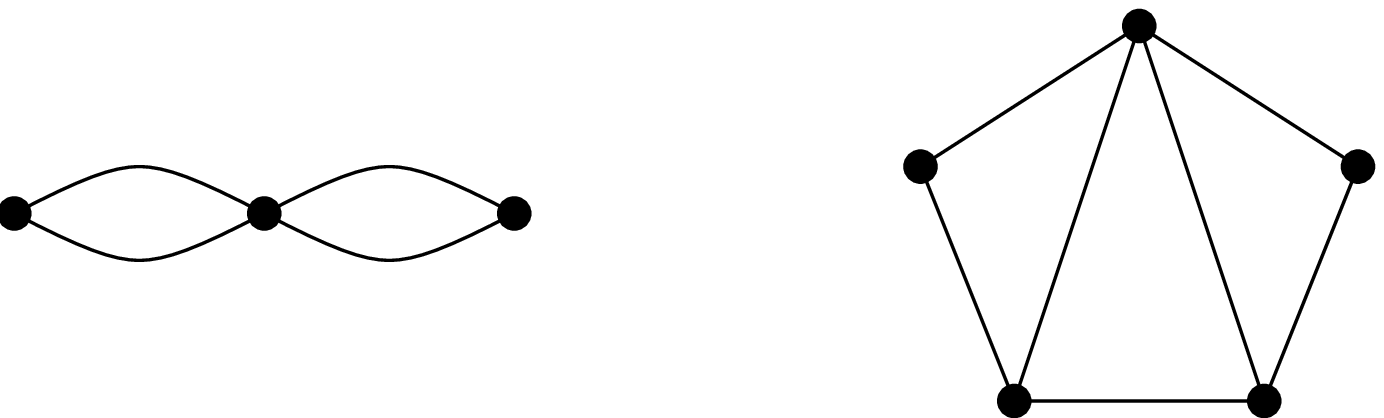}}
\def\fourtad{\epsfxsize=1cm\epsfbox{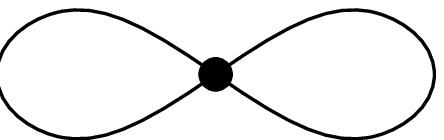}}
\def\schematic{\epsfxsize=12cm\epsfbox{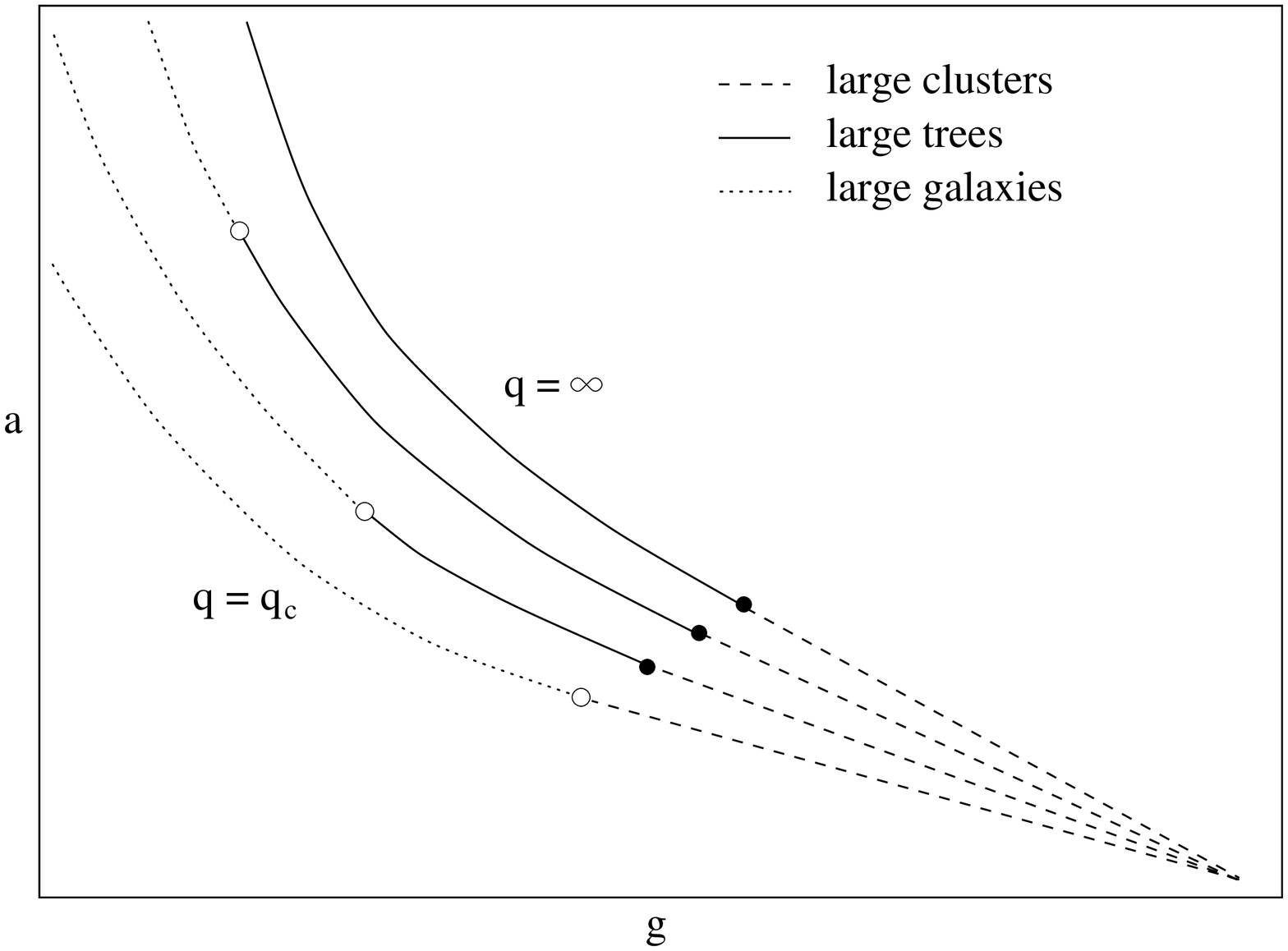}}
\def\nonlin{\ba\epsfxsize=1.5cm\epsfbox{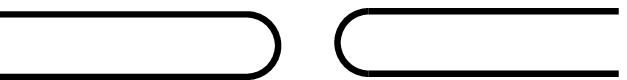}\ea}
\def\qplot{\epsfxsize=12cm\epsfbox{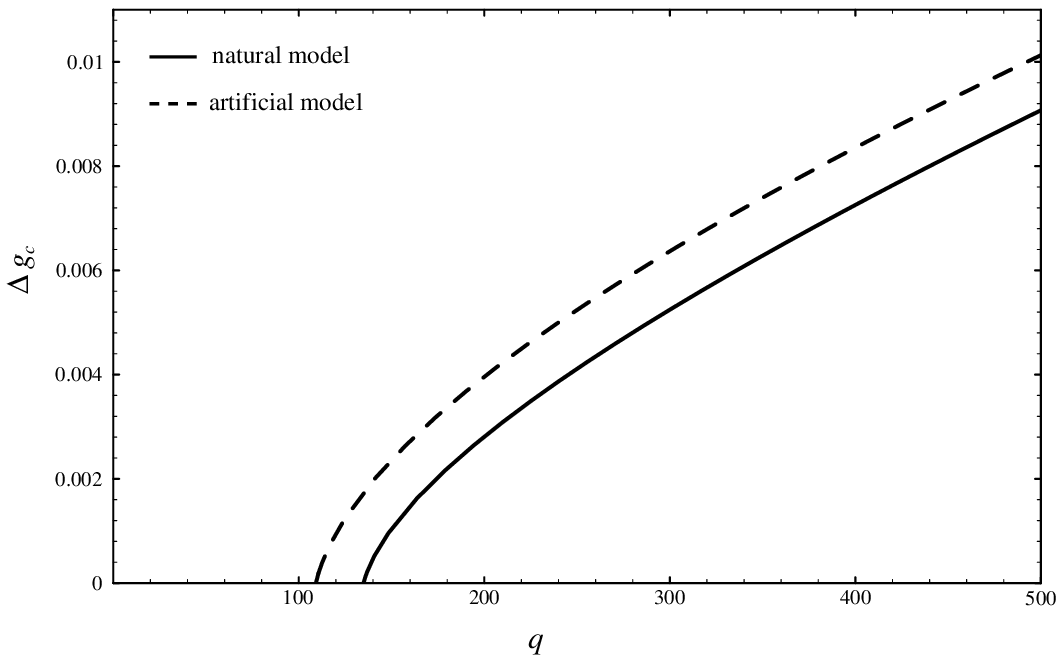}}

\newcommand{\appsection}[1]{\addtocounter{section}{1}\section*{#1}}

\newcommand{\D}{\Delta}
\newcommand{\oh}{\frac{1}{2}}
\newcommand{\ot}{\frac{1}{3}}
\newcommand{\of}{\frac{1}{4}}

\newcommand{\Tr}{{\rm Tr}\,}
\newcommand{\La}{\Lambda}
\newcommand{\la}{\lambda}
\newcommand{\be}{\begin{equation}}
\newcommand{\ee}{\end{equation}}
\newcommand{\gstr}{\gamma_{\rm str}}

\begin{document}
\vspace*{-3.3cm}
\vbox{
\begin{flushright}
NBI-HE-94-28 \\
May 1994
\end{flushright}
\title{From Trees to Galaxies:\\
The Potts Model on a Random Surface}
\author{Mark Wexler \\ \\
      \small {\it Niels Bohr Institute, Blegdamsvej 17}\\
      \small {\it 2100 Copenhagen \O, Denmark}\\ 
      \small {\it wexler@nbivax.nbi.dk}}
\date{}
\maketitle
\begin{abstract}
The matrix model of random surfaces with $c=\infty$ has
recently been solved and found to be identical to a random
surface coupled to a $q$-states Potts model with $q=\infty$.
The mean field-like solution exhibits a novel type of tree
structure.  The natural question is, down to which---if
any---finite values of $c$ and $q$ does this behavior persist?
In this work we develop, for the Potts model, an expansion
in the fluctuations about the $q=\infty$ mean field solution.
In the lowest---cubic---non-trivial order in this expansion
the corrections to mean field theory can be given a nice
interpretation in terms of structures (trees and ``galaxies'')
of spin clusters.  When $q$ drops below a finite $q_c$, the
galaxies overwhelm the trees at all temperatures, thus
suppressing mean field behavior.  Thereafter the phase
diagram resembles that of the Ising model, $q=2$.
\end{abstract}}

\section{Introduction} 

The infinite-state Potts model on random spherical
surface can be solved exactly \cite{Me2,Me3} by resumming all
orders of its low temperature expansion \cite{Me1}.  The
problem and its solution are of interest for the following
reasons.
\begin{itemize}
\item
It can be shown to be equivalent to conformally invariant
models with $c = \infty$ (which can be realized, for example,
by coupling multiple Ising models to the surface).  Thus the
solution opens a window onto the mysterious $c>1$ regime,
hitherto inaccessible via either Liouville theory or matrix
models.
\item
This equivalence to conformal matter holds in spite of a
well known fact that on a fixed two-dimensional lattice
a Potts model with more that four states suffers a
first order---rather than a continuous---phase transition.
Coupling an $\infty$-state Potts model to
a random surface changes the order of the transition
from first to third; in the case of $q<4$
the order of the transition correspondingly changes
from second to third.
\item
The reason why the transition for $q=\infty$ is of third
order is that it is no longer a spin-ordering transition;
rather it is a tree-growing transition, where the nodes
of the trees are spin clusters (equivalently, ``baby
universes'').  The transition in the random surface
Potts model from spin-ordering (low $q$) to tree-growing
(high $q$) behavior is a novel one, and deserves attention
in its own right.
\end{itemize}

Some recent work has given support to the mean field
theory at $q=c=\infty$.  Ambj\o rn, Durhuus and
J\'onsson \cite{ADJ,Durhuus} (see also \cite{ABC})
have shown that if a random
surface model has susceptibility exponent $\gstr$, its
``polymerized'' version has $\bar{\gamma}_{\rm str}
=\gstr/(\gstr-1)$,
in agreement with the mean field result $\gstr = \ot$.
Their polymerized surfaces of baby universes
are suggestive of the trees of clusters found in the
mean field model.\footnote{Although baby universes
(geometrical structures) are not {\it a priori} identical
to clusters (matter structures), the two seem to fuse
at large $q$ or $c$.  Note that the trees in \cite{ADJ}
are a result of the $N\to\infty$ limit, while the
trees in \cite{Me2,Me3} are a result of $q,c\to\infty$.}
Harris and Wheater \cite{HarrisWheater}
have studied multiple Ising models,
and in the limit $c\to\infty$ have identified trees as
dominant, and found
$\gstr=\oh$.\footnote{Their trees are made of
{\it vertices}, not {\it clusters}, and thus form
a subset of trees-of-clusters.  There is no contradiction:
to find clusters at the vertices of the trees, one
has to take $c\to\infty$ and $T\to0$ simultaneously,
keeping $e^{-2/T}c$ fixed.}
A recent numerical study \cite{MonteCarlo} of multiple
$q=2, 3,$ and $4$ Potts models shows that spikiness of
the surfaces increases with increasing central charge,
and that $\gstr$ may well approach $\ot$.
Finally, a current numerical study of the Potts model
for various values of $q$ and for surfaces with tadpoles
and self-energies \cite{Teaser} is confirming the scaling of the
critical temperature $T_c^{-1} = \oh \log q + {\cal O}(1)$
with large $q$, and, more significantly, details of
cluster geometry predicted by the mean field theory.

For convenience, let us introduce the concept of a
{\it skeleton graph}: given a triangulated surface with some spin
configuration, its skeleton is a graph that represents
the spin clusters as simple vertices, and whose edges are
the links between different spin clusters.\footnote{So
be definition, no vertex of a skeleton may be connected
to itself.}
Each loop of skeleton graph introduces a factor $1/q$
(or $1/c$ for conformal models).
Thus at $q=\infty$ (or at $c=\infty$),
the only skeletons that are present are trees.
When we move down to large but finite $q$, the corrections
introduce loops; and while the leading terms are universal,
the corrections in $1/q$ are not---their details depend
on the details of the matter model.
This is a familiar picture in $1/D$ expansions.  For instance
in the scalar $\lambda\phi^4$ theory, the upper critical
dimension is 4: below this, fluctuations or ``loops''
become essential, and mean field theory breaks down.

It would be very interesting to know if something similar
happens for random surfaces.  One's curiousity is heightened
by the fact that Liouville theory seems to be silent on the
matter.  To investigate the large-but-finite $q$ region,
we develop a systematic expansion of the model (in its
equivalent one-matrix form) about the $q=\infty$ solution.
A priori, we would expect that the first non-trivial order
in this expansion to be quadratic, as in many similar cases.
In this order we sum all the skeletons that have one or
no loops, in other words the $q^0$ and $q^{-1}$ terms in
the partition function.  But it is not hard to show that
this $q^{-1}$ correction is insignificant, in that it does not
modify the critical behavior of the model.

The next, cubic, order of our approximation is much richer.
Here we include all skeletons whose clusters have three
or fewer external legs that lie on loops.  We therefore
have terms with arbitrarily many loops, meaning arbitrarily
high powers of $q^{-1}$.  A new structure makes its appearance:
the clusters are organized into $\phi^3$ graphs of spherical
topology.  These ``surfaces'' of clusters will be called
{\it galaxies}.  The trees are still present: any number
of them may grow out of any cluster.  On top of that,
a typical skeleton graph includes several galaxies,
linked also into a tree.

This complicated system shows complicated critical behavior.
For $q=\infty$, there is a magnetized low temperature phase where the
clusters are large, and there are few of them: essentially
pure gravity.  At some point that phase becomes unstable,
and the tree of clusters blows up.  Above that, the clusters
are small, but the tree is large.

At large but finite $q$, the above picture of critical behavior
is unchanged, except for the appearance of a new phase
at high temperature,
that of large galaxies.  At low temperature we still have
the pure gravity phase of large clusters; this still goes
into an intermediate phase of large trees.  But now, at an even higher
temperature (which is infinite when $q$ is infinite, and
which decreases with $q$), the area of the galaxies---the
number of clusters they contain---diverges.  In this phase,
the loops become essential: without loops there could be no
galaxies.  The phase of large galaxies is also akin to
pure gravity.

The tree phase becomes narrower and narrower as $q$ decreases,
until at some $q_c$, it disappears altogether: one goes
from the magnetized phase directly into the phase
of large galaxies.  Although mean field theory is modified
at any finite $q$, it is completely destroyed below $q_c$.
For $q \le q_c$ the phase diagram is altogether reasonable,
in that it resembles the known phase diagrams of the $q<4$
models: there are two pure gravity phases, at low and at
high temperature, separated by a multicritical point.

This paper is organized as follows.  Section 2 presents the
multi-matrix version of the Potts model, its low temperature
expansion, its effective one-matrix versions, and its
$q=\infty$ solution.
In section 3 we introduce an expansion about the $q=\infty$
solutions and its truncated versions, and discuss its
geometrical implications.
In section 5 we
analyze the critical behavior of the simplest nontrivial
truncation.

\section{Definition of the model}

The $q$-state Potts model on a spherical random surface
will be defined using a matrix model (see \cite{GinspargMoore}
for a review):
\be
\label{basicpotts}
F_q(g,\zeta) = \frac{1}{q} \log \frac{1}{f_0}
\int {\cal D} \phi_1 \cdots {\cal D} \phi_q \:
\exp -N \, \Tr \left( \frac{1}{2} \sum_{ij}
\phi_i G_{ij} \phi_j + g \sum_i \phi_i^3 \right)
\ee
where the $\phi_i$ are $N \times N$ hermitian matrices
and all matrix integrals are to be understood as being
divided by $N^2$ and taken in the limit $N \to \infty$.
The matrix $G$ defines
the matter that is to be coupled to the fluctuating
surface; in the case of the Potts model it is
\be
G = \left(
\begin{array}{cccc}
1 & \zeta & \cdots & \zeta \\
\zeta & 1 & \ddots & \vdots \\
\vdots & \ddots & \ddots & \zeta \\
\zeta & \cdots & \zeta & 1
\end{array}
\right)^{-1}
= \: \left(
\begin{array}{cccc}
a_q & b_q & \cdots & b_q \\
b_q & a_q & \ddots & \vdots \\
\vdots & \ddots & \ddots & b_q \\
b_q & \cdots & b_q & a_q
\end{array} \right)
\ee
\be
\label{aqbq}
a_q = \frac{1 + (q-2) \zeta}{(1-\zeta)(1 + (q-1)\zeta)},\qquad
b_q = \frac{-\zeta}         {(1-\zeta)(1 + (q-1)\zeta)}
\ee
The coupling constant $\zeta$ is related to the matter
temperature by $\zeta = e^{-2/T}$.
Finally, the normalization $f_0$ is defined by requiring
$F_q(0,\zeta) = 1$.

The partition function $F_q$ can be expanded in powers of
$\zeta$.  This is the so-called low temperature or cluster
expansion, as in order $\zeta^n$ we have a sum over all connected
configurations of clusters with $n$ inter-cluster links;
as well as over the {\it internal} geometry of each cluster.
Defining $\D = q - 1$,
the first four orders of the expansion are
\begin{eqnarray}
\label{lte}
\lefteqn{F_q(g,c) = \pi_0 +
\oh \D \pi_1^2 \zeta + 
\left[\oh \D^2 \pi_1^2 (\pi_2 + \pi_{11}) +
   \of \D \pi_2^2 \right] \zeta^2 } \\
& & \hspace{1.2cm} \gz
    \hspace{.8cm} \gaa
    \hspace{1.7cm} \gaab
    \hspace{1.7cm} \gbb \nonumber \\
& & \mbox{\vspace*{-1cm}} \nonumber \\
& & \mbox{} + \left[\oh \D^3 \pi_1^2 (\pi_2 + \pi_{11})^2 +
\frac{1}{6} \D^3 \pi_1^3 (\pi_3 + \pi_{12} + \pi_{111})
+ \frac{1}{6} \D(\D-1) \pi_2^3 \right. \nonumber \\
& & \vspace*{-1cm}
    \hspace{1.5cm} \gaabb
    \hspace{2.8cm} \gaaab
    \hspace{2.9cm} \gbbb \nonumber \\
& & \qquad\left. \mbox{} + \oh \D^2 \pi_1 \pi_2 \left(\pi_3
+ \frac{\pi_{12}}{3}\right) +
\frac{1}{24} \D \pi_3^2\right] \zeta^3 + \cdots \nonumber \\
& & \hspace{2.4cm} \gabc
    \hspace{2.1cm} \gcc
\end{eqnarray}
The picture underneath each term is the corresponding
skeleton graph---see the Introduction.
The various coefficients $\pi_p$
($p$ is a partition of an integer $n$, where $n$
is the cluster's external legs)
are functions of the cosmological constant $g$ that sum
over the internal geometries of the clusters.

To define and to calculate the $\pi$'s, one introduces
an external matrix integral
\be
\label{kgndef}
\Pi(g,\La) = -\frac{1}{2N} \Tr\La^2 -
\log \frac{1}{p_0} \int {\cal D}\phi \:
\exp -N \, \Tr \left( \La\phi + \oh \phi^2 +
g \phi^3 \right)
\ee
This integral has been calculated by Kazakov and by
Gross and Newman.
If we expand it in powers of $\La$, we get
\begin{eqnarray}
\label{kgnexp}
\Pi(g,\La) & = & \pi_0 + \frac{\pi_1}{N} \Tr \La +
\oh \left[\frac{\pi_2}{N} \Tr \La^2 +
\frac{\pi_{11}}{N^2} (\Tr \La)^2 \right] \nonumber \\
& & \qquad + \left[ \frac{\pi_3}{N} \Tr \La^3 +
\frac{\pi_{12}}{N^2} \Tr \La \Tr \La^2 +
\frac{\pi_{111}}{N^3} (\Tr \La)^3 \right] + \cdots
\end{eqnarray}
The coefficients $\pi_p(g)$ are the ones we encounter
in the low temperature expansion (\ref{lte}).
In $n$-th order, we find $\pi_p$ for all partitions
$p$ of $n$.  The coefficient $\pi_{n_1 n_2 \cdots}$
is a connected Green's function of the one-matrix
model $\Pi(g,0)$ with $n_1 + n_2 + \cdots\:$ external
legs, on each of which we place the matrix $\La$;
the graphs can be twisted in various ways, which
gives rise to the different combinations of traces.
For example, $\pi_2$ sums graphs such as
$$
\notwist \quad \leadsto \quad \Tr \La^2
$$
while $\pi_{11}$ sums graphs such as
$$
\onetwist \quad \leadsto \quad (\Tr\La)^2
$$
The untwisted coefficients $\pi_n$ are just
the $n$-point connected BIPZ Green's function \cite{BIPZ},
while the twisted coefficients have not been discussed
before in the literature.  They can all be calculated
exactly, and the first few are given in
Appendix A.

The complicated linear combinations
of the coefficients $\pi_p$ appearing
in the low temperature expansion (\ref{lte}) are due
to the large $N$ limit of our model.  It will be seen
below that there is a way to organize these linear
combinations that will clarify their nature and simplify
their calculation.

Here we briefly summarize the solution of the Potts
model in the $q\to\infty$ limit.
As it can easily be seen that the coefficient of $\zeta^n$
in the partition function is an $n$-th degree polynomial
in $\D$, the proper scaling for the temperature is
$\zeta \sim 1/\D$.  Defining $\zeta = a/\D$ and keeping $a$
fixed\footnote{This is the most interesting way of taking
the $q\to\infty$ limit.  If we keep $\zeta$ fixed instead,
for example, we will find only one of the several phases
of the model.}
while letting $q \to \infty$, it was found that
only skeleton graphs that are trees contribute to
$F_\infty(g,a)$.  These can be resummed to give
\begin{eqnarray}
\label{qinfint}
F_\infty(g,a) & = & \lim_{\theta\to\infty} \frac{1}{\theta}
\log \frac{1}{f_0} \int_{-\infty}^{\infty} dx \:
\exp -\theta \left[ \frac{x^2}{2a} - \Pi(g, x) \right]\\
\label{qinfsol}
& = & \Pi(g, y) - \frac{y^2}{2a},\qquad
\Pi(g,y) \equiv \Pi(g,y1)
\end{eqnarray}
where the function $y(g,a)$ is a saddle point of
$F_\infty$:
\be
\label{qinfsaddle}
y = a \frac{\partial}{\partial y} \Pi(g,y) = a \Pi_1(g,y)
\ee
The general features of the $q=\infty$ model's critical
behavior have already been described in the Introduction;
see \cite{Me3} for details.\footnote{This limit can
be called a mean field solution because it can be
arrived at by (an {\it a priori} inexact) procedure
of substituting $\phi_i \phi_j \to \Phi \phi_i$ in
the multi-matrix model (\ref{basicpotts}) and solving
self-consistently for $\Phi$; additionally, one
has to take the (not {\it a priori} obvious) ansatz
that $\Phi$ is a multiple of unity, as in Wilson's
mean field solution of QCD.}

Kazakov \cite{Kazakov} has pointed out
that the random surface Potts model
can also be formulated as a one-matrix model with the action
\be
\label{natural}
S_n = \D \left[\frac{1}{2\hat{a}} \Tr\phi^2 -
N \left(1 + \frac{1}{\D}\right)
\Pi\left((1 - a/\D)^{3/2} g, \phi\right) \right],
\quad \hat{a} = \frac{a}{1 - a/\D}
\ee
The simple derivation is given in Appendix B.
This effective action will be called the ``natural'' one.
For the purposes of the present calculation, it is useful to
introduce a slightly different, ``artificial'' effective action
\be
\label{artificial}
S_a = \D \left[-\rho\, \Tr\phi + \frac{1}{2a} \Tr\phi^2 -
N \Pi(g,\phi) \right]
\ee
The difference between these two effective models, their relative
merits and drawbacks, are fully explained in
Appendix B.

In going from the multi-matrix model (\ref{basicpotts})
to the effective one-matrix models (\ref{natural}) and
(\ref{artificial}), we trade the complexity of having
several matrices for the complexity of the new one-matrix
potential---see eq.\ (\ref{kgnexp}).  But what is really
happening is that the vertices of the effective models
are playing a completely different role than the vertices
of the original model.  In the original multi-matrix model,
the vertices represented points on the discretized
worldsheet.  In the effective models, the vertices---which
are complicated functions of the original cosmological
constant $g$---actually represent entire spin clusters.
This is possible because a twisted cluster is equivalent,
from the large-$N$ fatgraph point of view, to a nonlinear
vertex
$$
\onetwist \quad \Longleftrightarrow \quad \nonlin
$$
such as $(\Tr\La)^2$.  This is why all possible nonlinear
terms appear in the expansion (\ref{kgnexp}) of the
potential $\Pi(g,\La)$; or, seen from the other direction,
why clusters of every possible twist appear in the low
temperature expansion (\ref{lte}).\footnote{The
procedure that led from multi-matrix model to the effective
one-matrix model can be applied to any---not only Potts---matter.
Only the Potts model, however, is symmetric with respect
to {\it any} relabeling of matter states; that is why its
effective model is a {\it one}-matrix model.  Effective
cluster models for other types of matter are multi-matrix
models.}

One could rederive the $q,\D \to \infty$ limit
in a quick and dirty way by noticing that
when the model (\ref{natural}) is reduced to an eigenvalue
problem, the charge of the eigenvalues will scale as
$\D^{-1}$, and therefore the width of the eigenvalue
spectrum will scale as $\D^{-1/2}$.  At $\D=\infty$, the
eigenvalue density will be a delta function---precisely
the content of equation (\ref{qinfint}).

\section{Fluctuations about the mean field}

So at $q=\infty$ the eigenvalue density of the effective
models (\ref{natural}) and (\ref{artificial}) collapses
to a delta function.
When $q$ is large but finite, the eigenvalues
spread into a narrowly peaked distribution.  The main idea
of this work is to investigate the shape of that distribution
by expanding the effective models about the scalar solution
\be
\label{phipsi}
\phi = y 1 + \psi
\ee
in powers of $\psi$.  One type of critical point will result
from singularities of the equations for the eigenvalue
distribution of $\psi$.
Once an approximate distribution is found in this way, we will
plug it back into the full effective model to determine critical
behavior;
another type of critical point will result from singularities
of the potential $\Pi$ for a given $\psi$.
For simplicity, the following calculations will be shown only for
the artificial model (\ref{artificial}); the calculations for the
natural model (\ref{natural}) differ in trivial ways, and their
result will be given at the end.

The action (\ref{artificial}) now reads
\be
\label{intermact}
S_a = \D \left[ N\left(-\rho y + \frac{y^2}{2a}\right) +
\left(-\rho + \frac{y}{a}\right) \Tr \psi
+ \frac{1}{2a} \Tr\psi^2
-N \Pi(g,y+\psi) \right]
\ee
The reader is reminded \cite{Me3}
that the $q=\infty$ partition function
$F_\infty$ is a sum over free trees, and its saddle point $y(g,a)$
is a sum over {\it planted} trees: trees with a marked vertex
of order one.  Explicitly,
\be
\label{yexpl}
y(g,a) = \pi_1 a + \pi_1 (\pi_2 + \pi_{11}) a^2 +
\left[\pi_1 (\pi_2 + \pi_{11})^2 +
\oh \pi_1^2 (\pi_3 + \pi_{12} + \pi_{111})\right] + \cdots
\ee
This low temperature expansion of $y$ can be represented
graphically as
$$
\eqy
$$
We are led to consider the expansion of the external matrix integral
about a scalar:
\begin{eqnarray}
\label{expwithy}
\lefteqn{
\Pi(g,y + \psi) = \Pi_0(g,y)
+ \frac{\Pi_1(g,y)}{N} \Tr \psi}\\
& & \quad \mbox{} + \oh \left[
\frac{\Pi_2(g,y)}{N} \Tr \psi^2
+ \frac{\Pi_{11}(g,y)}{N^2} (\Tr \psi)^2 \right] \nonumber \\
& & \quad \mbox{} + \frac{1}{6} \left[
\frac{\Pi_3(g,y)}{N} \Tr \psi^3
+ \frac{\Pi_{12}(g,y)}{N^2} \Tr \psi \Tr \psi^2
+ \frac{\Pi_{111}(g,y)}{N^3} (\Tr\psi)^3 \right]
+ \cdots \nonumber
\end{eqnarray}
The vertex functions $\Pi_p$ can be calculated explicitly, and are
given in Appendix A.
To understand what they are counting,
we can expand several of them in powers of $y$:
\begin{eqnarray}
\Pi_0(g,y) & = & \pi_0 + \pi_1 y + \oh (\pi_2 + \pi_{11}) y^2
+ \frac{1}{6} (\pi_3 + \pi_{12} + \pi_{111}) y^3 + \cdots\nonumber \\
\Pi_2(g,y) & = & \pi_2 + \left(\pi_3 + \frac{\pi_{12}}{3}\right) y
+ \oh \left(\pi_4 + \frac{\pi_{13}}{2} + \frac{\pi_{22}}{3} +
\frac{\pi_{112}}{6} \right) y^2 + \cdots\nonumber \\
\label{cubexpy}
\Pi_3(g,y) & = & \pi_3 + \left(\pi_4 + \frac{\pi_{13}}{4}
\right) y + \cdots,\: {\rm etc.}
\end{eqnarray}
The linear combinations of $\pi_p$'s in $\Pi_0, \Pi_2, \Pi_3,
\ldots$ are precisely those that occur in the low temperature
expansion (\ref{lte}).  The $n$-th coefficient of $\Pi_0$
counts clusters that have $n$ external legs, each of which is
connected to a tree, {\it i.e.}, none of its legs reconnect.
By contrast,
the $n$-th coefficient of $\Pi_2$ counts clusters that have
two legs that lie on a loop (and will eventually reconnect),
and $n$ other legs that are connected to trees; and so on.
Thus a cluster with three external legs, for example, may
appear in the low temperature expansion as $\pi_3$, $\pi_3
+ \pi_{12}/3$, or $\pi_3 + \pi_{12} + \pi_{111}$, depending
on whether it has 3, 2, or 0 legs that lie on loops.

The expansion of action (\ref{intermact}) in powers of
$\psi$ is therefore\footnote{Two terms cancel due to
(\ref{qinfsol}).}
\begin{eqnarray}
\label{artfull}
\frac{S_a}{\D} & = & -N \left(\Pi_0 - \frac{y^2}{2a} + \rho y\right)
-\rho\,\Tr\psi + \oh (a^{-1} - \Pi_2) \Tr \psi^2 -
\frac{\Pi_{11}}{2N} (\Tr\psi)^2 \nonumber \\
& & \qquad\mbox{} - \frac{\Pi_3}{6} \Tr \psi^3
- \frac{\Pi_{12}}{6 N} \Tr \psi \, \Tr \psi^2
- \frac{\Pi_{111}}{6 N^2} (\Tr\psi)^3 + \cdots
\end{eqnarray}
According to the above discussion, the vertex $\Pi_3\,\Tr\psi^3$
is really an infinite sum of vertices
(\ref{cubexpy}), and can be graphically
represented as
\be
\label{vertfigone}
\Pi_3(g,y)\,\Tr\psi^3 \quad \leadsto \quad
\begin{array}{c}\vertnotwist\end{array}
\ee
while a non-linear vertex such as $\Pi_{12}\,\Tr\psi\,\Tr\psi^2$
may be represented as
\be
\label{vertfigtwo}
\Pi_{12}(g,y)\,\Tr\psi\,\Tr\psi^2 \quad \leadsto \quad
\begin{array}{c}\vertonetwist\end{array}
\ee
The model (\ref{artfull}) is an expansion about the tree-like,
mean field solution $\phi = y$.  The geometrical meaning of
this statement is reflected in the above pictures: trees are
automatically included in the vertices of the new matrix model.
The graphs of this matrix model can be pictured as a dense
network (in fact, a spherical surface) of clusters, from the
vertices of which sprout trees.

To proceed farther with the matrix
model (\ref{artfull}) we are obliged to truncate
the potential $\Pi(g,y+\psi)$.
The simplest possibility is to truncate
after the quadratic term.
We take $\rho=0$ in this case,
and immediately evaluate the partition
function\footnote{The last term in (\ref{trunctwo})
is there to cancel cluster self-con\-nec\-tions; it should have
been put in by hand as a constant in the action.}
\be
\label{trunctwo}
F = \Pi_0(y) - \frac{y^2}{2a}
- \frac{1}{2\D} \log[1 - a \Pi_2(y)] - \frac{a \Pi_2(y)}{2\D}
\ee
The new $1/\D$ terms in the partition function just count the
one-loop contributions to the low temperature expansion
(\ref{lte}).
It is clear that they cannot effect the critical
behavior of the model: a large tree with one loop is
indistinguishable from just a large tree.  This can
be demonstrated by noticing that the only way in which
the $1/\D$ terms can effect the critical behavior of the model
is through a divergence of the logarithm;
but, since $\Pi_2 < \partial_y^2
\Pi_0 = \Pi_2 + \Pi_{11}$ (see Appendix A),
and $1 - a \partial_y^2
\Pi_0 \ge 0$ in the physical region of the $q=\infty$ model
(see \cite{Me2}), this is impossible.

We are therefore obliged to keep higher order terms in the
potential $\Pi(g,y+\psi)$.
The meaning of a truncation after $\psi^n$ terms,
for $n \ge 3$, is as follows.
Consider a cluster with $\la$ links leading to other clusters.
Of those links, $\la_t$ are to trees and $\la_\ell = \la - \la_t$
are to loops; meaning that if one of the $\la_t$ tree links is
cut, the skeleton will be disconnected.  For instance, the cluster
indicated by the arrow
$$
\exvert
$$
has $\la_t = 2$ and $\la_\ell = 3$.  So: if we truncate the
action (\ref{artfull}) after the $\psi^n$ terms, we discard
any skeleton that has a cluster with $\la_\ell > n$.  Note
that $\la_t$ does not matter: as discussed above, each vertex
of (\ref{artfull}) automatically includes arbitrary numbers
of trees.

We now consider the model (\ref{artfull}) in the {\it cubic
approximation}: we discard all terms of order $\psi^4$ or
higher.  In this approximation the
clusters---which are themselves spherical $\varphi^3$
graphs---are arranged into spherical $\varphi^3$ graphs
that also have trees of clusters
growing from their vertices (given by
powers of $y$: see (\ref{vertfigone}), (\ref{vertfigtwo})).
This structure will be
called a ``galaxy.''  A typical graph in the cubic
model has several galaxies, with the nonlinear terms
$(\Tr\phi)^2$, $\Tr\phi\,\Tr\phi^2$, and $(\Tr\phi)^3$
connecting the galaxies into a tree structure (not to
be confused with the trees that grow out of the vertices
inside the galaxies).  The cubic approximation differs
radically from the quadratic one in that it does not simply
sum over graphs with one more loop and with one higher
order in $\D^{-1}$, but rather it sums over a large
class of graphs with an arbitrary number of loops.
In fact, it sums over all graphs, provided that no
cluster has more than three legs lying on loops;
for example, the following are simple skeleton graphs
that are {\it not} included in the cubic approximation:
$$
\notcubic
$$

The solution of the cubic model will proceed along the
same lines as the simpler model solved in \cite{Indians}.
First off, we set
\be
\rho = -\frac{\Pi_3}{2a} - \frac{\Pi_{12}}{6a}
\ee
to cancel the cluster self-connections.  Then we follow
the standard procedure, reducing the problem to eigenvalues
and introducing a continuous approximation to the spectrum
$\la(x)$.  Using 
procedure that is exact in the $N\to\infty$ limit
we introduce the two moments
\be
c = \left\langle\frac{1}{N}\Tr\psi\right\rangle = \int dx\,\la(x),
\qquad
d = \left\langle\frac{1}{N}\Tr\psi^2\right\rangle=\int dx\,\la^2(x)
\ee
as parameters, for which will we solve self-consistently
at the end.  The saddle point equation for the action reads
\begin{eqnarray}
\label{saddlepoint}
\frac{\delta S}{\delta\la} & = & q + m \la(x) + 3 p \la^2(x)
- \frac{1}{2} \int' \frac{dy}{\la(x)-\la(y)} = 0 \\
-\frac{p}{\D} & = & \frac{\Pi_3}{6} \\
-\frac{m}{\D} & = & -\frac{1}{a} + \Pi_2 + \frac{c \Pi_{12}}{3} \\
-\frac{q}{\D} & = & \rho +  c \Pi_{11} +
\frac{d\Pi_{12}}{6} + \frac{c^2 \Pi_{111}}{2}
\end{eqnarray}
keeping in mind that the $\Pi_p$'s are (complicated) functions
of $g$ and $y(g,a)$.
The eigenvalue density for this problem is
\be
\label{evaldens}
u(\la) = \frac{1}{\pi} (m + \sigma + 3 p \la)
\sqrt{\frac{4}{m + 2\sigma} - \left(\la - \frac{\sigma}{3p}\right)^2}
\ee
where the variable $\sigma$ satisfies the equation
\be
\label{sigma-eq}
18 p^2 + \sigma(m+\sigma)(m+2\sigma) + 3 p q(m+2\sigma) = 0
\ee
Using the eigenvalue density (\ref{evaldens}), the self-consistency
conditions on $c$ and $d$ become
\begin{eqnarray}
\label{c-eq}
c & = & \frac{3p}{(m+2\sigma)^2} + \frac{\sigma}{3p} \\
\label{d-eq}
d & = & \frac{m+4\sigma}{(m+2\sigma)^2} +
\left(\frac{\sigma}{3p}\right)^2
\end{eqnarray}

The reader is reminded that the eigenvalue density (\ref{evaldens})
is for the $\psi$ matrix.  Since $\sigma$ scales as $\D^0$, for
large $\D$ the eigenvalue density for $\phi$ in the cubic
approximation is centered on the $q=\infty$ value $y$ with
a displacement $\sigma/3p \sim {\cal O}(\D^{-1})$.  The width
of the distribution is $\sqrt{4/(m+2\sigma)} \sim {\cal O}
(\D^{-1/2})$, as previously advertised.

\section{Critical behavior}

The cubic approximation to the Potts model is expected
to have three critical phases: the phase of {\it large
clusters}, the phase of {\it large trees} of clusters,
and the phase of {\it large galaxies} of clusters.

First, the regime of {\it large clusters}.
This is the magnetized, low temperature phase.
As we have an approximate form for the eigenvalue density
(\ref{evaldens}), we can plug it back into the full action
(\ref{artificial}) and look for the singularities of the
potential $\Pi(g,\phi)$.  The reader is reminded
\cite{Me3} that for $q=\infty$, the eigenvalue density
is a delta function at $y(g,a)$; the potential in that
case can easily be reduced to the ``standard'' form
$\phi^2/2 + G \phi^3$, with $G = (1 - 12 g y)^{-3/4} g$.
At the critical point
\be
\label{qinfcrit}
G=G_c=1/\sqrt{108\sqrt{3}}
\ee
the area of the clusters (but not their number) diverges.
The same thing happens in the case of the spread out density
(\ref{evaldens}), though the computation is somewhat more
difficult: we are looking for the critical point of the
external matrix model (\ref{kgndef}).
The result can be found in \cite{GrossNewman}.
The only dependence on the external matrix is through
the moments
\be
\label{defmoments}
\sigma_k = \int d\la \, u(\la) (x - \la)^{-k/2}
\ee
where the variable $x$ satisfies the implicit equation
\be
\label{eqgn}
x = \frac{1}{12g} - \sqrt{3g} \sigma_1(x)
\ee
For the eigenvalue density, these moments can be
calculated explicitly (as functions of $x$):
\begin{eqnarray}
\lefteqn{\sigma_k = \left(x - y - \frac{\sigma}{3p}\right)^{-k/2}
\left[\mbox{}_2 F_1\!\left(\oh+\frac{k}{4},\frac{k}{4};2;
\frac{4}{(1+2\sigma) (x - \sigma/3p)^2}
\right)\right.} \\
& & \left.\qquad\mbox{} + \frac{3kp}{2(1+2\sigma)^2
(x - \sigma/3p)}
\,\mbox{}_2 F_1\!\left(\oh+\frac{k}{4},1+\frac{k}{4};3;
\frac{4}{(1+2\sigma) (x - \sigma/3p)^2}
\right)\right]
\nonumber
\end{eqnarray}
The criticality condition is the point where the
$x$ equation (\ref{eqgn}) ceases to have solutions.
Algebraically this can be expressed as
\be
\label{stupidcrit}
3 g \sigma_3^2 = 4
\ee
It can easily be checked that when $q,\D\to\infty$,
and consequently $\sigma/3p,4/(m+2\sigma) \to 0$,
these new criticality conditions (\ref{eqgn}) and
(\ref{stupidcrit}) reduce to (\ref{qinfcrit}).
It will be seen in the phase diagrams that as
$q$ approaches infinity, the curve of large clusters
will approach the corresponding $q=\infty$ curve,
given by solving eqs.\ (\ref{qinfcrit}) and 
(\ref{qinfsaddle}) simultaneously.

In the large cluster phase matter is manifestly
magnetized, and is therefore decoupled from the
geometry; so this is a phase of pure gravity,
with $\gstr = -\oh$.  Another way of saying the
same thing: in this phase the number of clusters
is small, and few clusters is the same as one
cluster, as far as critical behavior is concerned.

Second, the phase of {\it large trees} of clusters.
There are, in principle, two ways of getting there.
At $q=\infty$, the trees become large when the saddle
point equation for $y(g,a)$ (\ref{qinfsaddle})
ceases to have a solution \cite{Me3}; in other words, when
\be
\label{qinfsaddlecrit}
a \frac{\partial^2}{\partial y^2} \Pi(g,y)
= a \left[\Pi_2(g,y) + \Pi_{11}(g,y)\right] = 1
\ee
The other way to get large trees is to make the
tree of {\it galaxies} (of clusters) blow up.
This is not exactly the same as a large tree of
clusters at $q=\infty$, but very nearly the same,
as in general the galaxies at the vertices of this
tree will be small; and as far as critical behavior
is concerned, a small galaxy is no different from
a galaxy of size one, in other words, a single
cluster.  This intuition is supported by the fact
that both large tree phases have $\gstr = +\oh$;
and as $q\to\infty$, the curves of the two phases
approach each other.

To calculate the location of the second large tree
phase, we can look for the point in the cubic model
(\ref{saddlepoint}) where the string susceptibility
$\partial^2 F/\partial p^2$ diverges \cite{Indians}.
Equivalently, and computationally simpler, we can
look for the divergence of the
``pseudo-susceptibility''\footnote{$d'$ or $\sigma'$
would do just as well.}
\be
\label{defpseudo}
c' \equiv \frac{\partial c}{\partial p} \to \infty
\ee
This is most simply done by differentiating the
eqs.\ (\ref{sigma-eq}--\ref{d-eq}) with respect to
$p$, and solving the resulting linear system for $c'$,
$d'$, and $\sigma'$.
As the result is unwieldy, and in any case can easily
be derived, it will not be given here.  The same procedure,
if applied to the simple quartic model in \cite{Indians},
reproduces the result for the large tree phase given in that
work.

In the large tree phase one has an ensemble of clusters
of different size.  As one moves toward the magnetized
phase, clusters of higher and higher area begin to be
included.  At the multicritical point, some
moments of the cluster area diverge, a sign 
of a continuous spin ordering phase
transition.\footnote{For the Ising model
on a fixed two-dimensional lattice, for example, the
second moment of the cluster size distribution is the
lowest one to diverge.\cite{Fisher}}  The susceptibility
exponent in the bulk of the tree phase is
$\gstr = \oh$, the generic exponent for tree growth.
In the tree phase, and only in the tree phase, the
clusters are probably identical to the geometric structures
known as ``baby universes'' \cite{MonteCarlo}.
Therefore the transition to large trees from high
temperature can be interpreted as a magnetization
of baby universes (but not of the whole surface);
and the transition to trees from low temperature can
be interpreted as a break-up of the magnetized surface
into magnetized (but in different directions) baby
universes \cite{Teaser}.

Finally, the cubic model should also have a phase
of {\it large galaxies}.
This occurs simply at the normal, BIPZ, critical point 
of the model (\ref{saddlepoint}).  This is simply given
by
\be
\label{largegalaxies}
\Gamma = \frac{p}{(m - 12 p q)^{3/4}} = \Gamma_c =
\frac{1}{\sqrt{108 \sqrt{3}}}
\ee
This phase is radically different from anything that
exists at $q=\infty$; it is in this phase that loops
become important.  Interestingly, it bears some resemblance
to what {\it must} be the case for small $q$: the spin
clusters at high temperature do not form trees,
but complicated planar networks with many loops.

The large planar networks of clusters in the galaxy
phase resemble the configurations of a low $q$ model
at high temperature.  As we go to higher temperature,
moreover, the clusters get smaller, as matter fluctuations
are once again decoupled from the geometry.  The large
galaxy phase is therefore pure gravity, once again.
The susceptibility exponent is $\gstr = -\oh$, the
standard pure gravity value for nonlinear models in
this phase.

We are now ready to study the complicated
transcendental equations of the cubic approximation.
The procedure is as follows.  First, fix the three
adjustable parameters: $\D = q-1$, $g$, and $a$.
Then solve the saddlepoint equation (\ref{qinfsaddle})
for $y(g,a)$; this fixes all coefficients in the
cubic action (\ref{artfull}).  Next, solve the
cubic equation (\ref{c-eq}) for $c$ as a function
of $\sigma$, plug into (\ref{sigma-eq}) and solve
numerically for $\sigma$, taking the root closest
to zero.  Finally, check the criticality conditions:
(\ref{eqgn}) and (\ref{stupidcrit}) for the phase
of large clusters; (\ref{defpseudo}) for large
trees; or (\ref{largegalaxies}) for large galaxies.

The following is a schematic, not-to-scale sketch
of the resulting critical curves.
$$
\schematic
$$
If drawn to scale, the curves are rather close to
each other and hard to distinguish by eye.
As $q\to\infty$, the curves reassuringly approach
the exact $q=\infty$ critical curve, where the
transition from magnetized to tree phases occurs
at $g = 1/\sqrt{288}, a = 1$.  As we move
away from $q=\infty$, the new critical regime of
large galaxies appears.  As $q$ gets lower, the
phases of large galaxies and large clusters
approach each other, eating away at the mean field,
large tree phase.

The susceptibility exponent $\gstr$ is $-\oh$ in the
magnetized and galaxy phases, as discussed above,
and $+\oh$ in the tree phase.  At the two multicritical
points $\gstr = \ot$.  Further, we find $\alpha=-1$
at both multicritical points, as shown in Appendix C.
The equality of the exponents at the two multicritical
points is not a coincidence: the model at one of the
points can be mapped onto the model at the other---see
the appendix for details.

At $q=q_c$ ($q_c \approx 133$ for the natural model,
and $q_c \approx 108$ for the artificial model), the
phase of large trees ceases to exist.  Thereafter,
the magnetized phase of large clusters goes
directly into a phase of large galaxies.
The mean field behavior is entirely eliminated.
Let $g_{c1}(q)$ be the multicritical point between
phases of large galaxies and large trees, and
$g_{c2}(q)$ the multicritical point between 
the phases of large trees and large clusters.
$\D g_c = g_{c2} - g_{c1}$ is a measure of
``how much of mean field theory is left'' at a
given value of $q$.  Here it is plotted for both
the natural and the artificial models.
\qplot

For $q \le q_c$, the two pure gravity phases
(magnetized and large galaxies) are joined directly
onto each other, with a multicritical point in
the middle.  This phase diagram bears a striking
resemblance to the known phase diagram for, say,
$q=2$ \cite{KazakovIsing}.

\section{Discussion}

The results obtained for $q_c$ cannot be considered
as anything more than order-of-magnitude estimates.
The real result of this calculation is the discovery of
the mechanism for the break-down
of mean field theory: the divergence
of galaxies.  The resulting critical behavior interpolates
smoothly between the somewhat odd phase diagram at $q=c=\infty$
and the well-known phase diagram for $q \le 4, c \le 1$.
Another result of this calculation is the likelihood
that $q_c$ is {\it finite}, that mean field behavior
holds for some finite Potts
models (with modifications, as has been shown).

The two natural questions at this stage are: is the
present result merely an artifact of the truncation of the
action (\ref{artfull}), or are the higher-order terms
essential?  And does a similar picture hold for conformally
symmetric models, which, after all, reduce to the same
mean field theory at $c=\infty$?

The first question should not be too hard to answer.
It seems that we would not find any qualitatively new
behavior by including the higher-order terms, which
would simply add other types of vertices to the galaxies
but would not modify the essential structure.
Mean field theory should always be destroyed by the
divergence of the galaxies.
It is true that quartic or higher models will add
new multicritical points; but as we only have the three
parameters $q, g,$ and $a$ to tune, we should not be
able to reach them.
To repeat the present calculation in the quartic approximation
should be somewhat tedious, but not too difficult.

The second question is more difficult to answer, and the
author does not have any suggestions on how to deal with
it.
The immediate difficulty is that, although effective
cluster models such as (\ref{natural}) and (\ref{artificial})
can be introduced for any matter model on the surface,
they are multi-matrix rather than one-matrix models.
Still, they closely resemble the $c=\infty$ model, so
perhaps it is the right way to start.

\vspace{1cm}
The author thanks the Niels Bohr Institute for
its hospitality, and Jan Ambj\o rn and Gudmar Thorleifsson
for stimulating discussions.  Marc Potters provided many
of the ideas in the early stage of this work.

\appendix

\appsection{Appendix~\thesection}
\label{App:PiDetails}

Here we give expressions for the first few Green's functions
$\pi_p(g)$ in the expansion (\ref{kgnexp}) of the external
matrix integral $\Pi_(g,\La)$, defined in (\ref{kgndef}).
Using Gross and Newman's notation, the external matrix
integral is
\begin{eqnarray}
\Pi(g,\Lambda) & = &
-\frac{1}{2N^2} \sum_{a,b} \log(\mu_a 
+ \mu_b) - \frac{1}{6g} (\sigma_{-2} - x) + 
\frac{2}{\sqrt{27g}}\sigma_{-3} \nonumber \\
& & \qquad\mbox{} + \sigma_1 \sigma_{-1} + \sqrt{\frac{g}{48}} 
\sigma_1^3 - \frac{1}{108g^3} - \frac{1}{4} \log 3g
\label{app:GrossNewmanIntegral}
\end{eqnarray}
where the $\la_a$ are eigenvalues of $\La$, $\mu_a = \sqrt{x -
\la_a}$, $\sigma_k = \frac{1}{N} \sum_a (x - \la_a)^{-k/2}$,
and the variable $x$ satisfies the implicit equation
(\ref{eqgn}).

First we expand $x$ in powers of $\La$:
\begin{eqnarray}
\lefteqn{x = x_0 - \frac{\sqrt{3g}}{\sqrt{3g} - 2 x_0^{3/2}}
\frac{\Tr\La}{N}} \\
& & \quad\mbox{} - \frac{9 g \left(\sqrt{3g} - 4 x_0^{3/2}\right)}
{4 x_0\left(\sqrt{3g} - 2 x_0^{3/2}\right)^3}
\frac{(\Tr\La)^2}{N^2} +
\frac{3\sqrt{3g}}{4 x_0 \left(\sqrt{3g} - 2 x_0^{3/2}\right)}
\frac{\Tr\La^2}{N} + \cdots \nonumber
\end{eqnarray}
where $x_0$ satisfies the cubic equation
\begin{eqnarray}
x_0 & = & \frac{1}{12g} - \sqrt{\frac{3g}{x_0}} \\
    & = & \frac{1}{18g} + \frac{1}{36}
		\left(\sqrt{3} \sin \frac{2\theta}{3} -
		\cos \frac{2\theta}{3} \right) \\
    &   & \qquad
		\tan \theta = \sqrt{\left(\frac{g_0}{g}\right)^4 - 1},
		\quad g_0 = \frac{1}{\sqrt{108 \sqrt{3}}} \nonumber \\
& = & \frac{1}{12g} - 6 g - 216 g^3 - \cdots \nonumber
\end{eqnarray}
Plugging the series for $x$ back into (\ref{app:GrossNewmanIntegral}),
after some algebra we find
\begin{eqnarray*}
\pi_1 & = & \sqrt{\frac{x_0}{3g}} + \frac{1}{4x_0} - \frac{1}{6g} \\
      & = & 3 g + 108 g^3 + 7776 g^5 + \cdots \\
\pi_2 & = & \frac{1}{2 \sqrt{3g x_0}} - \frac{1}{16 x_0^2} - 1 \\
	 & = & 27 g^2 + 1944 g^4 + \cdots \\
\pi_{11} & = & \frac{1}{16 x_0^2} \\
	    & = & 9 g^2 + 1296 g^4 + \cdots \\
\pi_3 & = & \frac{1}{4\sqrt{3g} x_0^{3/2}} - \frac{1}{16 x_0^3} \\
	 & = & 6 g + 540 g^3 + 58320 g^5 + \cdots \\
\pi_{12} & = & \frac{3}{16 x_0^3} \\
	    & = & 324 g^3 + 69984 g^5 + \cdots \\
\pi_{111} & = & \frac{3g}{8 x_0^3 (4x_0^3 - 3g)} +
			 \frac{\sqrt{3g}}{4 x_0^{3/2} (4x_0^3 - 3g)} \\
          & = & 7776 g^5 + \cdots
\end{eqnarray*}

The corresponding coefficients $\Pi_p$ in the expansion of
$\Pi(g,y + \La)$ (\ref{expwithy}) can be obtained in a similar
manner.  The above expressions for $\pi_p(g)$ go directly into
$\Pi_p(g,y)$ with the substitution $x_0 \to x_0 - y$, while $x_0$
itself now satisfies
\be
x_0 = \frac{1}{12g} - \sqrt{\frac{3g}{x_0 - y}}
\ee
and is given by
\begin{eqnarray}
x_0 & = & \frac{1}{18g} + \frac{y}{3} +
\frac{1 - 12 g y}{36} \left( \sqrt{3} \sin\frac{2\theta}{3} -
\cos\frac{2\theta}{3} \right) \\
& & \qquad \tan \theta = \sqrt{\left(\frac{g_0}{G}\right)^4 - 1},
\quad G = \frac{g}{(1 - 12 g y)^{3/4}} \nonumber
\end{eqnarray}

\appsection{Appendix~\thesection}
\label{App:OneMatrix}

Consider a model with an extra matrix $\phi$ and the
action
\be
\label{app:intermediateaction}
S' = \Tr \left( \phi \sum_i \phi_i - \frac{1}{2 b_q} \phi^2
+ \frac{a_q - b_q}{2} \sum_i \phi_i^2 + g \sum_i \phi_i^3
\right)
\ee
with $a_q, b_q$ given in (\ref{aqbq}).
On one hand,
the extra matrix can be integrated out to give exactly
the model (\ref{basicpotts}); one the other hand, the $\phi_i$
can be integrated out in (\ref{app:intermediateaction}) to give
a one-matrix model with the action
\be
\label{app:natural}
S_n = \D \left[\frac{1}{2\hat{a}} \Tr\phi^2 -
N \left(1 + \frac{1}{\D}\right)
\Pi\left((1 - a/\D)^{3/2} g, \phi\right) \right],
\quad \hat{a} = \frac{a}{1 - a/\D}
\ee
Whereas the vertices of the original multi-matrix model
(\ref{basicpotts}) are the points of a discretized
worldsheet, the vertices of the effective one-matrix
model (\ref{app:natural}) are spin clusters.

The partition function of the effective one-matrix model
(\ref{app:natural}) can be expanded in powers of $a$ (not
$\hat{a}$) to
reproduce the low temperature expansion given above
(\ref{lte}).  The reason that this works is that a vertex
like $(\Tr \phi)^2$, such as occurs in the action
(\ref{app:natural}), is equivalent---as far as large $N$
behavior is concerned---to a twisted cluster,
as explained in the text.
The correspondence between the two expansions
is still not completely trivial, because in
the low temperature expansion (\ref{lte})
\begin{itemize}
\item the propagator between clusters is $a$, rather than
$\hat{a}$;
\item a cluster is never linked to itself---by definition;
\item the coefficient of a skeleton graph is a polynomial
in $\D^{-1}$, whose leading term is $\D^{v - \ell - 1}$
(where $v$ and $\ell$ are the numbers of vertices and links
in the graph)\footnote{Assuming that the expansion (\ref{lte})
is rewritten in terms of $a$, rather than $c$}; if there
are no loops of length greater than two present, the polynomial
is just a monomial, but otherwise there are subleading terms
that arise due to the finite volume of target space---for
instance, the coefficient of\,\,
\Gbbb
\,\,\,\,is $\D^{-1}(1 - \D^{-1})$.
\end{itemize}
Evidently, the $1/\D$ terms in the action (\ref{app:natural})---the
correction to the coefficient of $\Pi$ and the renormalization
$g \to (1 - a/\D)^{3/2} g$---conspire to compensate for the above
three discrepancies.

In the calculation to be performed, the potential $\Pi$
in the effective one-matrix model (\ref{app:natural})
will be truncated after the cubic term (only the terms written
out in equation (\ref{kgnexp}) are included).
It is instructive to consider the effect of this truncation on
the low temperature expansion of the effective model.  Of course
the clusters will now have no more than three neighbors.  Furthermore,
consider, say, the coefficient of $a^2$:
\be
\label{app:tpovercomp}
\oh \pi_1^2 (\pi_2 + \pi_{11})
+ \frac{1}{4\D} \pi_2^2
- \frac{1}{12\D^2} \left(\pi_4 + \frac{\pi_{22}}{2}\right)
\ee
The first two terms reproduce the low temperature expansion
(\ref{lte}).  The last term, however, compensates for the tadpole
$$
\fourtad
$$
that would have been generated by the $\phi^4$ terms.
This over-compensation for tadpoles is somewhat undesirable.
We can consider an alternate, ``artificial'' action (as opposed to
the ``natural'' action (\ref{app:natural}))
\be
\label{app:artificial}
S_a = \D \left[-\rho\, \Tr\phi + \frac{1}{2a} \Tr\phi^2 -
N \Pi(g,\phi) \right]
\ee
with $\rho$ adjusted to cancel the simple tadpoles ({\it i.e.,}
the cluster self-con\-nec\-tions), in this case to $-(\pi_1 + 
\pi_3/2 + \pi_{12}/6) a$.  Strictly speaking, a constant
term should also be added to $S_a$ to cancel {\it all} the
cluster self-connections; but as this term will have no effect
whatsoever on the calculations here, it will be omitted.
The partition function of the
artificial model, when expanded in low temperature series,
reproduces the low temperature expansion (\ref{lte}) exactly,
{\it except} that the coefficient of each graph is precisely
$\D^{v-\ell-1}$, so that the coefficient of\,\,
\Gbbb
, for example, is $\D^{-1}$ rather than the correct
$\D^{-1}(1 - \D^{-1})$.
The artificial model (\ref{app:artificial})
does not know about the target space of the original Potts
model, and so it cannot reproduce its excluded volume effects.
On the positive side, the potential $\Pi$ in the artificial
action can be truncated at any point without introducing 
tadpole overcompensation terms such as (\ref{app:tpovercomp}).

To summarize, there are two effective one-matrix models
for the Potts model, the ``natural'' model (\ref{app:natural}) and
the closely related ``artificial'' model (\ref{app:artificial}).
If the potential $\Pi(\phi)$ in these models is truncated,
slight discrepancies (different for the two models) are 
introduced into their low temperature expansions, as compared
to the series (\ref{lte}).  These discrepancies are not
important when $q$ is large.  For instance, the critical
behavior of the natural and artificial models is identical,
up to slightly different critical parameter values.

\appsection{Appendix~\thesection}

Here we sketch the calculation of the critical exponent $\alpha$.
It is defined by the singular behavior of the partition function
$F \sim |g-g_c|^{2-\alpha}$, where we constrain the approach
to the multicritical points to lie on a critical curve.

At $q=\infty$, we can calculate $\alpha$ exactly for the multicritical
point between the large tree phase and the magnetized pure gravity
phase.  The partition function $F_\infty$ is given by eqs.\
(\ref{qinfsol}) and (\ref{qinfsaddle}).
The multicritical point occurs at $g_* = 1/\sqrt{288}, a_* = 1$.
We vary $g$, and constrain $a$ to lie on the critical curve.
When $g>g_*$, this is given by
\be
a_+(g) = \frac{1 - (g/g_0)^{4/3}}{1 - \sqrt{3} (g/g_0)^{2/3}
+ (g/g_0)^{4/3}}
\ee
where $g_0^2 = 1/108\sqrt{3}$ is the pure gravity critical point;
when $g<g_*$, the critical curve is given by the equation
(\ref{qinfsaddlecrit})
\be
a_- \frac{\partial^2}{\partial y^2} \Pi(g,y) =
a_- \left[\Pi_2(g,y) + \Pi_{11}(g,y)\right] = 1
\ee
in conjunction with eq.\ (\ref{qinfsol}).  The partition function
turns out to be analytic on either side of the multicritical point.
Putting $\D g=g - g_*$, and using the expressions given in
Appendix A, we find
\[
F_+(g) \equiv \lim_{g\searrow g_*} F_\infty(g,a_+(g))
= \left(\frac{7}{24} - \frac{1}{2} \log 2\right) + 96 \D g^2
- 2432 \sqrt{2} \D g^3 + \cdots
\]
\[
F_-(g) \equiv \lim_{g\nearrow g_*} F_\infty(g,a_-(g))
= \left(\frac{7}{24} - \frac{1}{2} \log 2\right) + 96 \D g^2
- 5760 \sqrt{2} \D g^3 + \cdots \\
\]
The discontinuity in the third derivatives means that the
transition is of third order, with $\alpha = -1$.
The corresponding transition from the magnetized to the
tree phase for $\infty < q < q_c$ is similar, and should
have the same exponents.

At the other multicritical point, the boundary between the
tree and galaxy phases, the geometric fluctuations of the
clusters (as given by the coefficients $\Pi_p$ in action
(\ref{artfull})) no longer matter; what drives the transition
is the {\it inter\/}-cluster dynamics of the galaxies, as
well as the inter-galaxy dynamics.  To calculate the critical
exponent we therefore may ignore the complexities of the
coefficients $\Pi_p(g,y)$ in the action, and consider any
simple non-linear matrix model which displays similar behavior.
As a first example, take the quartic model
\be
\label{nonlinquart}
S = \Tr \left(\oh \phi^2 + g \phi^4\right) +
\frac{g'}{N} (\Tr\phi^2)^2
\ee
that was solved in \cite{Indians}; the exponent $\alpha$
has never been given in the literature.\footnote{In the context
of the original work \cite{Indians}, this exponent is
meaningless, as there is no outright matter coupled to
the surface.}  The phase of ``large galaxies''
($\gstr = -\oh$) occurs on the curve $g' = g'_-(g)$ and the
phase of large trees ($\gstr = +\oh$) occurs on the curve
$g' = g'_+(g)$, with
\begin{eqnarray}
g'_+ & = & -g - \frac{1}{32} \left(1 + \sqrt{1 + 64 g}\right) \\
g'_- & = & -9 g - \frac{3}{4} \sqrt{-3g}
\end{eqnarray}
The multicritical point occurs at $g_* = -3/256, g'_* = -9/256$.
As above, we fix $g' = g'_\pm$ and find on either side of the
multicritical point
\[
F_+ = \left(\frac{29}{72} - \oh \log \frac{8}{3}\right)
- \frac{64}{27} \D g + \frac{8192}{27} \D g^2
+ \frac{8388608}{243} \D g^3 + \cdots
\]
\[
F_- = \left(\frac{29}{72} - \oh \log \frac{8}{3}\right)
- \frac{64}{27} \D g + \frac{8192}{27} \D g^2
- \frac{8388608}{243} \D g^3 + \cdots
\]
Therefore the transition is again of third order,
with $\alpha = -1$.

As another example, consider the nonlinear cubic model
\be
\label{nonlincubic}
S = \Tr \left(\oh \phi^2 + g \phi^3\right)
+ \frac{g'}{2N} (\Tr\phi)^2
\ee
which is closer to our effective model, eq.\ (\ref{artfull}).
Of course it has exactly the same critical behavior as the
quartic model (\ref{nonlinquart}), which can be
shown in an identical manner.  It is more instructive,
however, to consider the partition function of the model
(\ref{nonlincubic}) directly: it is simply given by
\be
F = \Pi(g,c g') - \frac{g' c^2}{2}
\ee
where $c$ is the one-point function, given by the
self-consistency condition
\be
c = \left\langle\frac{1}{N}\Tr\phi\right\rangle = \Pi_1(g,c g')
\ee
exactly analogous to (\ref{c-eq}).  If we put $y = c g'$,
we obtain precisely the partition function of the $q=\infty$
model (\ref{qinfsol}), along with its saddle point condition,
eq.\ (\ref{qinfsaddle}).  Therefore $\alpha = -1$, as calculated
above.

We have therefore mapped one of the multicritical points onto
the other.  A similar mapping holds for the full action
(\ref{artfull}), giving $\alpha = -1$ at the transition
between the tree and galaxy phases.  This mapping points
up the similarity of the magnetized phase
to the phase of large galaxies: in the former the individual
clusters blow up, while in the latter it is galaxies of
clusters that diverge.  Both phases are, essentially,
pure gravity.

\end{document}